\documentclass[times,11pt]{article}
\newcommand{\etal}{{\it et al.\ }}

\newcommand{\vs}{{\it vs.\ }}

\newcommand{\beq}{\begin{equation}
  \renewcommand{\int}{\intop\limits}
  \renewcommand{\oint}{\ointop\limits}}
\newcommand{\eeq}{\end{equation}}
\newcommand{\beqarr}{\par\begin{minipage}{11cm} \begin{eqnarray*}}
\newcommand{\eeqarr}{\end{eqnarray*} \end{minipage} \hfill 
   \stepcounter{equation}{\rm (\theequation)}\vspace{3mm}\linebreak}
\newcommand{\bdm}{\begin{displaymath}
  \renewcommand{\int}{\intop\limits}
  \renewcommand{\oint}{\ointop\limits}}
\newcommand{\edm}{\end{displaymath}}

\newcommand{\up}[1]{\ifmmode^{\rm #1}\else$^{\rm #1}$\fi}

\newcommand{\arcd}{\ifmmode^{\circ}\else$^{\circ}$\fi}
\newcommand{\arcm}{\ifmmode{'}\else$'$\fi}
\newcommand{\arcs}{\ifmmode{''}\else$''$\fi}

\newcounter{pagefrom}
\newcounter{pageto}
\newcounter{volume}
\newcounter{year}

\newenvironment{Titlepage}{
\vspace*{2cm}
  \begin{center}
}{
  \end{center}\par\vspace{3mm}
}

\newcommand{\Title}[1]{{\large\bf\boldmath #1 \\[3mm] {\footnotesize by} 
\\[3mm]}}

\newcommand{\Author}[2]{{\large\spaceskip 2pt plus 1pt minus 1pt #1}\\[3mm]
   {\small #2}\\[6mm]}

\newcommand{\Received}[1]{}

\newcommand{\Abstract}[2]{{\footnotesize\begin{center}ABSTRACT\end{center}
\vspace{1mm}\par#1\par
\noindent
{\bf Key words:~~}{\it #2}}}

\newcommand{\FigCap}[1]{\footnotesize\par\noindent Fig.\  %
  \refstepcounter{figure}\thefigure. #1\par}

\newcommand{\TabCap}[2]{\begin{center}\parbox[t]{#1}{\begin{center}
  \small {\spaceskip 2pt plus 1pt minus 1pt T a b l e}
  \refstepcounter{table}\thetable \\[2mm]
  \footnotesize #2 \end{center}}\end{center}}

\newcommand{\TableFont}{\footnotesize}

\newcommand{\MakeTable}[4]{\begin{table}[htb]\TabCap{#2}{#3}
  \begin{center} \TableFont \begin{tabular}{#1} #4 
  \end{tabular}\end{center}\end{table}}

\newcommand{\MakeTableSep}[4]{\begin{table}[p]\TabCap{#2}{#3}
  \begin{center} \TableFont \begin{tabular}{#1} #4 
  \end{tabular}\end{center}\end{table}}

{
\footnotesize \frenchspacing

\renewcommand{\and}{{\rm and }}
\section{{\rm REFERENCES}}
\sloppy \hyphenpenalty10000
\begin{list}{}{\leftmargin1cm\listparindent-1cm
\itemindent\listparindent\parsep0pt\itemsep0pt}}%
{\end{list}\vspace{2mm}}

\def\TYLDA{~}
\newlength{\DW}
\newcommand{\refitem}[5]{\item[]{#1} #2%
\def\REFARG{#3}\ifx\REFARG\TYLDA\else, {\it#3}\fi
\def\REFARG{#4}\ifx\REFARG\TYLDA\else, {\bf#4}\fi
\def\REFARG{#5}\ifx\REFARG\TYLDA\else, {#5}\fi.}

\usepackage{supertabular,lscape,epsfig}
\usepackage{amssymb}
\usepackage{amsmath}
\DeclareSymbolFont{ppa}{OT1}{ppl}{m}{it}
\DeclareMathSymbol{\vv}{\mathalpha}{ppa}{'166}
\begin{document}

\newcommand{\TabCapp}[2]{\begin{center}\parbox[t]{#1}{\centerline{
  \small {\spaceskip 2pt plus 1pt minus 1pt T a b l e}
  \refstepcounter{table}\thetable}
  \vskip2mm
  \centerline{\footnotesize #2}}
  \vskip3mm
\end{center}}

\newcommand{\TTabCap}[3]{\begin{center}\parbox[t]{#1}{\centerline{
  \small {\spaceskip 2pt plus 1pt minus 1pt T a b l e}
  \refstepcounter{table}\thetable}
  \vskip2mm
  \centerline{\footnotesize #2}
  \centerline{\footnotesize #3}}
  \vskip1mm
\end{center}}

\newcommand{\MakeTableSepp}[4]{\begin{table}[p]\TabCapp{#2}{#3}
  \begin{center} \TableFont \begin{tabular}{#1} #4
  \end{tabular}\end{center}\end{table}}

\newcommand{\MakeTableee}[4]{\begin{table}[htb]\TabCapp{#2}{#3}
  \begin{center} \TableFont \begin{tabular}{#1} #4
  \end{tabular}\vspace*{-7mm}\end{center}\end{table}}

\newcommand{\MakeTablee}[5]{\begin{table}[htb]\TTabCap{#2}{#3}{#4}
  \begin{center} \TableFont \begin{tabular}{#1} #5
  \end{tabular}\end{center}\end{table}}

\newcommand{\FigurePs}[7]{\begin{figure}[htb]\vspace{#1}
\includegraphics{#4}
\FigCap{#2\label{#3}}
\end{figure}}

\newfont{\bb}{ptmbi8t at 12pt}
\newfont{\bbb}{cmbxti10}
\newfont{\bbbb}{cmbxti10 at 9pt}
\newcommand{\uprule}{\rule{0pt}{2.5ex}}
\newcommand{\douprule}{\rule[-2ex]{0pt}{4.5ex}}
\newcommand{\dorule}{\rule[-2ex]{0pt}{2ex}}
\def\thefootnote{\fnsymbol{footnote}}

\begin{Titlepage}
\Title{The All Sky Automated Survey. The Catalog of Variable Stars.
IV.~18$^{\rm\bf h}$--24$^{\rm\bf h}$ Quarter of the Southern Hemisphere}
\Author{G.~~P~o~j~m~a~{\'n}~s~k~i}{Warsaw University Observatory,
Al~Ujazdowskie~4, 00-478~Warszawa, Poland\\
e-mail:gp@astrouw.edu.pl}
\Author{Gracjan Maciejewski}{Toru\'n Centre for Astronomy, N.
Copernicus University, ul.~Gagarina~11, 87-100~Toru\'n, Poland\\
e-mail:Gracjan.Maciejewski@astri.uni.torun.pl}
\end{Titlepage}

\Abstract{ In this paper we present the fourth part of the
photometric data from the $9\arcd \times 9\arcd$ ASAS camera
monitoring the whole southern hemisphere in $V$-band. Preliminary
list (based on observations obtained since
January 2001) of variable stars located between 
RA $18^{\rm h}$ and $24^{\rm h}$ is released. 
10311 stars brighter than $V$=15 were found to be
variable (1641 eclipsing, 1116 regularly pulsating, 938 Mira and
6616 other stars). Light curves have been classified using the
automated algorithm taking into account periods, amplitudes, fourier 
coefficients of the light curves, 2MASS
colors and IRAS infrared fluxes. Basic photometric properties are
presented in the tables and some examples of thumbnail light
curves are printed for reference.
All photometric data are  available over the INTERNET at
{\it http://www.astrouw.edu.pl/\~{}gp/asas/asas.html} or {\it http://archive.princeton.edu/\~{}asas}}{Catalogs --Stars: variables: general -- Surveys}

\section{Introduction}
The All Sky Automated Survey (ASAS, Pojma\'nski 1997) is a
photometric CCD sky survey aiming at continuous monitoring of
bright objects of the southern hemisphere. It was initiated by
ideas of Paczy\'nski (1997) of performing many interesting
scientific programs with small automated instruments.

The ASAS system is located in Las Campanas Observatory (operated
by the Carnegie Institution of Washington) and consists of four
independent instruments installed together in a small automated
enclosure. Every wide-field camera ($9\arcd \times 9\arcd$) is
composed of f200/2.8 telephoto lenses, 2K$\times$2K CCD equipped
with standard filter and a parallactic mount (Pojma\'nski 2001).
This configuration allows to obtain brightness measurements of all
sources brighter than limiting magnitude $V\sim~14$ ($I\sim~13$).

Variability analysis was performed on the $V$-band data as soon as
reasonable amount of data had been collected. We have already
presented preliminary catalogs of variable stars in the $0^h-6^h$
(Pojma\'nski 2002), $6^h-12^h$ (Pojma\'nski 2003) and $12^h-18^h$
(Pojma\'nski and Maciejewski 2004) quarters the southern hemisphere. This paper
contains the fourth part of the analyzed data - variable stars
located in the fields centered between $18^h$ and $24^h$ of right
ascension and declination $\delta < 0\arcd$.

\section{Observations and Data Reduction}

The data reduction pipe-line used to process ASAS data was
described in details by Pojma\'nski (1997). Its most important
feature is simultaneous photometry made through five apertures (2
to 6 pixels in diameter). Each aperture data is processed
separately,  so one can use data obtained with the smallest one
for the faint ($V > 12$) stars and with the largest one for the
bright ($V < 9$) objects. Substantial differences between large-
and small-aperture magnitudes indicate close neighbors in which
case large apertures should be avoided.

The astrometric calibration of CCD frames was based on the ACT
(Urban \etal 1998) catalog. The typical positional accuracy is
better than 0.2 pixels ( $<3$~arc sec).

The zero-point offset of photometry is based on the Tycho
(Perryman {\em et al.} 1997) data. A few hundred Tycho stars
usually located in each $9\arcd\times 9\arcd$ field, were used for
precise offset calibration. Due to effects caused by non-perfect
flat-fielding, lack of color terms in transformation and blending
of the stars, systematic errors as large as several tens of
magnitude could be observed for particular stars. Differential
accuracy is much better, reaching 0.01 for bright stars.

\section{Variability Search}

The search for variables in the fourth quarter of the southern sky
was done in the middle of 2004. Data used for variability
classification covered three years (since 2001) of observation.

Variability analysis was similar to that performed on the ASAS-2
data (Pojma\'nski 2000). Only stars with highest magnitude
dispersion (top 5 \%) and having large number of deviating
observations were subject to Analysis Of Variance test
(Schwarzenberg-Czerny 1989). Objects with AOV statistics value
larger than 10., and those that have passed long-term variability
tests: variance analysis in variable-length bins and trend
analysis (average number of consecutive observations showing the
same direction of brightness change) were visually inspected.

\section{Variability Classification}

Variability  classification was described in details in 
previous parts of this catalog (Pojma{\'n}ski 2002, 2003, 2004), 
so here we list briefly  only basic steps of the algorithm:
\begin{itemize}
\item{All stars are searched for 2MASS and IRAS counterparts, providing 
$J,H,K$ for all stars (10\% of stars are subject to some abiguity due to 
blending) and infrared fluxes for some of them.}
\item{
Using dedicated filter sensitive to deviations from the averaged light curve,
they are divided into strictly periodic and less regular ones.
}
\item{Strictly periodic variables are analyzed with an automatic
classification algorithm (that uses period, amplitude, Fourier 
coefficients of the light curve, $H-K$ and  $J-H$ colors and infrared fluxes).
Results are sorted according 
to the variability type and verified visually.}
\item{Less periodic light curves are tested for location in and 
$H-K$ \vs $J-H$ plane. If they fall within area of the late-type irregular 
or semi-regular stars, they are simply assigned MISC type. All other cases 
are inspected individually.
}
\end{itemize}

There is a group of DCEP or CW stars with quite periodic behavior but 
apparently variable amplitude. We have added "/SR" suffix to their 
variability type. 

\section{The Catalog}

We have selected 10311 variable stars in the fourth quarter of the
Southern Hemisphere.

For each star the following data are provided: ASAS identification $ID$
(coded from the star's $\alpha_{2000}$ and $\delta_{2000}$ in the
form: $hhmmss-ddmm.m$), period $P$ in days (or characteristic time
scale of variation for irregular objects), $T_0$ -  epoch of minimum
(for eclipsing) or maximum (for pulsating) brightness, $V_{max}$ -
brightness at maximum, $\Delta V$ - amplitude of variation, $Type$ - one
of the predefined classes: $DSCT$, $RRC$, $RRAB$, $DCEP_{FU}$, $DCEP_{FO}$,
$CW$, $ACV$, $BCEP$, $M$ and $MISC$. GCVS cross-identification,
$J$, $J-H$ and $H-K$ data taken from 2MASS catalog are also provided.

Stars classified as MISC contain mostly semi-regular and irregular variables 
as well as many objects excluded from other classes
after visual inspection (e.g. Novae). 
 861  objects other than MISC have multiple classification.
For 362 cases this is exclusively due to $EC/ESD$ or $ED/ESD$ confusion,
but for 452 other this is a more
serious $EC/RRC$ or $EC/DSCT$ double classification.

\tabcolsep 5pt
\MakeTable{|l|r|l|r|}{8cm}{\label{tabvar}
Number of various types of variable stars detected in the fourth quarter of the southern sky by ASAS-3 $V$ camera.}{
\hline
\multicolumn{1}{|c|}{Type} & \multicolumn{1}{c|}{Count} & \multicolumn{1}{c|}{Type} & \multicolumn{1}{c|}{Count}\\
\hline
$DCEP_{FU}$ & 118   & $DSCT$&  256 \\
$DCEP_{FO}$ &  23   & $EC$  &  943 \\
$CW$        &  127   & $ED$  & 256 \\
$ACV$       &  34   & $ESD$ & 442 \\
$BCEP$      &  14   & $M$   & 938 \\
$RRAB$      & 341   & $MISC$& 6616 \\
$RRC$       & 203   &       &  \\
\hline
}

Search for GCVS (Kholopov, \etal 1985) variables revealed about 2454 possible
matches within 3 arc minute radius.

Table 1. summarizes our classification effort and
Table 2. contains a compact version of the
catalog. Only four  columns are
listed  for each star: identification $ID$, $P$, $V$, and $\Delta V$.
Column $ID$ also contains some
flags - ":" if classification was uncertain, "?" if multiple classes
were assigned (objects were grouped in the table according to the highest rank
assignment), "v" if SIMBAD lists a star to be variable.

Appendix shows exemplary light curves. Only $ID$ is given for each.
For periodic variables
phase in the range ($-0.1$ - $2.1$) is plotted along the $x$-axis, while for
Mira's and miscellaneous stars -  HJD in the range (2451800-2453400).
Larger ticks on $y$-axis always mark
1 magnitude intervals and vertical span is never smaller than 1 mag.

The full catalog of variable stars observed by the ASAS system,
containing more classification details, as well as complete data for the
light curves, is available over the INTERNET:\\
\centerline{\it http://www.astrouw.edu.pl/\~{}gp/asas/asas.html}
 or
\centerline{\it http://archive.pinceton.edu/\~{}asas}

\section{Conclusions}

In this paper we present the fourth part 
of the preliminary catalog of variable stars in the Southern Hemisphere.
10,311 variable stars located between $18^{\rm h}$ and $24^{\rm h}$ 
are listed, increasing the total number of variables detected by ASAS to
over 35,000.

The ASAS catalog is still incomplete and our future efforts will 
concentrate on a) repeating variability search using substantially larger 
samples of data, b) searching for low-amplitude variables and c) searching 
for variable stars in the equatorial area of the sky.

\section{Acknowledgments}
This project was made possible by a generous gift from Mr. William
Golden to Dr. Bohdan Paczy{\'n}ski, and funds from Princeton University.  It is a
great pleasure to thank Dr. B. Paczy{\'n}ski for his initiative, interest,
valuable discussions, and the funding of this project.

I am indebted to the OGLE collaboration (Udalski, Kubiak, Szyma{\'n}ski 1997)
for the use of facilities of the
Warsaw telescope at LCO, for their permanent support and maintenance of the
ASAS instrumentation, and to The Observatories of the
Carnegie Institution of Washington for providing
the excellent site for the observations.

This research has made use of the SIMBAD database,
operated at CDS, Strasbourg, France and
of the NASA/ IPAC Infrared Science Archive, which is operated by the Jet Propulsion Laboratory, California Institute of Technology, under contract with the National Aeronautics and Space Administration.

This work was supported by the KBN 2P03D02024 grant.

\vspace{1.5cm}
\begin{center}
References
\end{center}

\noindent
\begin{itemize}
\leftmargin 0pt
\itemsep -5pt
\parsep -5pt
\refitem{Kholopov, P.N., \etal}{1985}{~}{~}{General Catalog of Variable
Stars, The Fourth Edition, Nauka, Moscow}
\refitem{Paczy\'nski, B.}{1997}{~}{~}{``The Future of Massive
Variability Searches'', in {\it Proceedings of 12th IAP
Colloquium}: ``Variable Stars and the Astrophysical Returns of
Microlensing Searches'', Paris (Ed. R. Ferlet), p.~357}

\refitem{Perryman, M.A.C. \etal}{ 1997}{ Astron. Astroph}{ 323}{
L49}

\refitem{Pojma\'nski, G.}{1997}{Acta Astron.}{47}{467}

\refitem{Pojma\'nski, G.}{2000}{Acta Astron.}{50}{177}

\refitem{Pojma{\'n}ski, G.}{2001}{~}{~}{
``The All Sky Automated Survey (ASAS-3) System - Its Operation and Preliminary 
Data'' ini: Small Telescope Astronomy on Global Scales,
{\it ASP Conference Series} Vol. 246, IAU Colloquium 183. Edited by
Bohdan Paczynski, Wen-Pin Chen, and Claudia Lemme. San Francisco:
Astronomical Society of the Pacific, p. 53}{}{}

\refitem{Pojma\'nski, G.}{2002}{Acta Astron.}{52}{397}

\refitem{Pojma\'nski, G.}{2003}{Acta Astron.}{53}{341}

\refitem{Pojma\'nski, G., Maciejewski, G.}{2004}{Acta
Astron.}{54}{153}

\refitem{Udalski, A. Kubiak, M., and Szyma\'nski, M.}{1997}{Acta
Astr.}{47}{319}

\refitem{Schwarzenberg-Czerny, A.}{1996}{Astrophys. J.}{460}{L107}

\refitem{Urban,S.E., Corbin T.E., Wycoff,
G.L.}{1998}{AJ}{115,1709}{2161}
\end{itemize}

\tabcolsep 3pt
\input{vartab.tex}

\textheight 24cm
\begin{figure}[p]
\vglue-3mm
\centerline{\bf Appendix}
\vskip1mm
\centerline{\bf ASAS Atlas of Variable Stars. ${\bf 18^{\bf h}{-}24^{\bf h}}$
Quarter of the Southern Hemisphere}

\centerline{\small Only several light curves of each type are printed. Full Atlas
 is
available over the {\sc Internet}:}

\centerline{\it http://www.astrouw.edu.pl/\~{}gp/asas/appendix18.ps.gz}
\vskip20mm
\centerline{Stars classified as EC}
\vskip3.5cm
\centerline{Stars classified as ESD}
\vskip3.5cm
\centerline{Stars classified as ED}
\vskip3.5cm
\centerline{Stars classified as DSCT}
\vskip-11.7cm
\centerline{\includegraphics[bb=30 70 400 495, width=13cm]{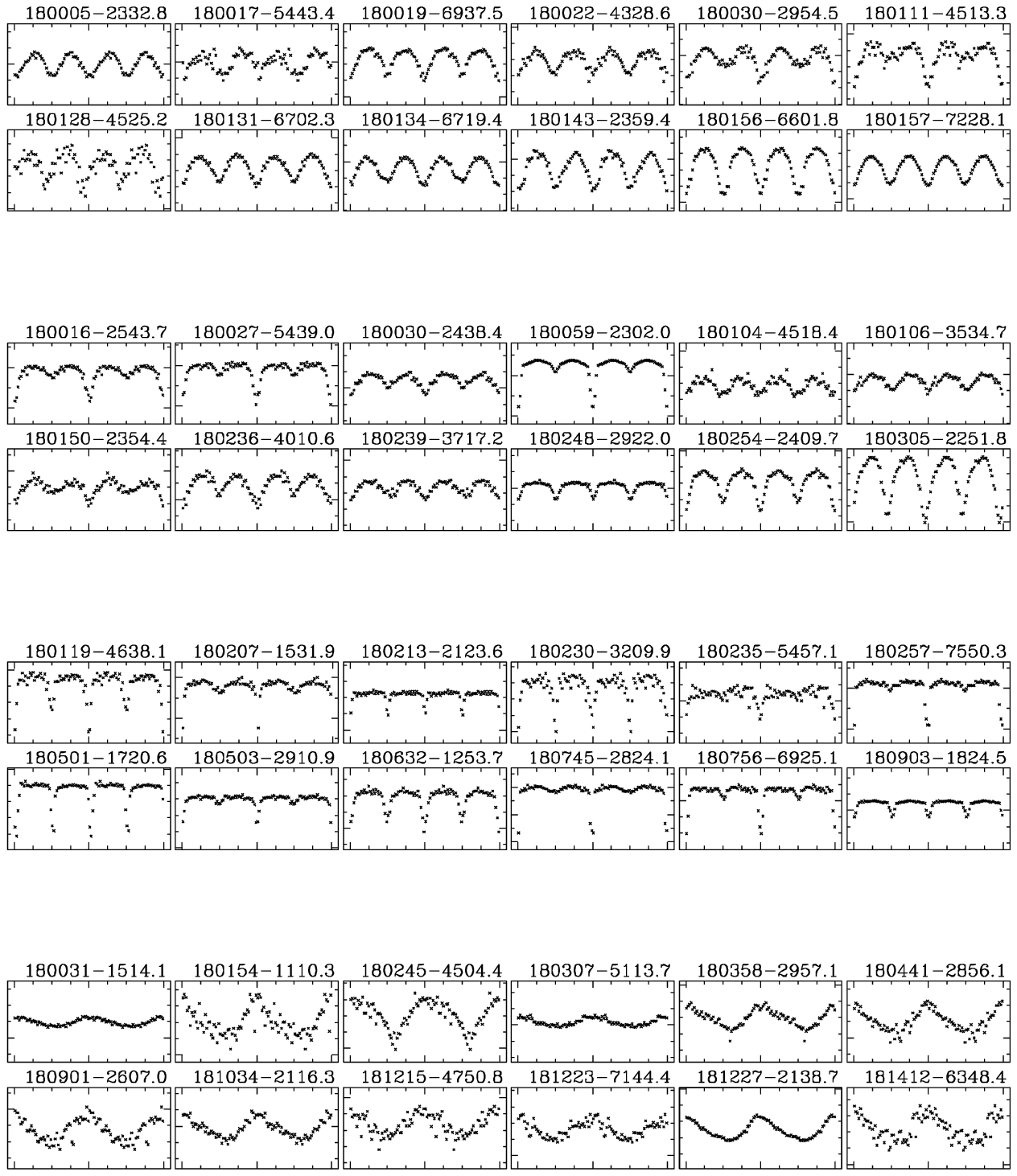}}
\end{figure}
\begin{figure}[p]
\vskip-2mm
\centerline{Stars classified as BCEP}
\vskip3.5cm
\centerline{Stars classified as ACV}
\vskip3.5cm
\centerline{Stars classified as RRC}
\vskip3.5cm
\centerline{Stars classified as RRAB}
\vskip3.5cm
\centerline{Stars classified as DCEP-FU}
\vskip-15.7cm
\centerline{\includegraphics[bb=30 30 405 610, width=13cm]{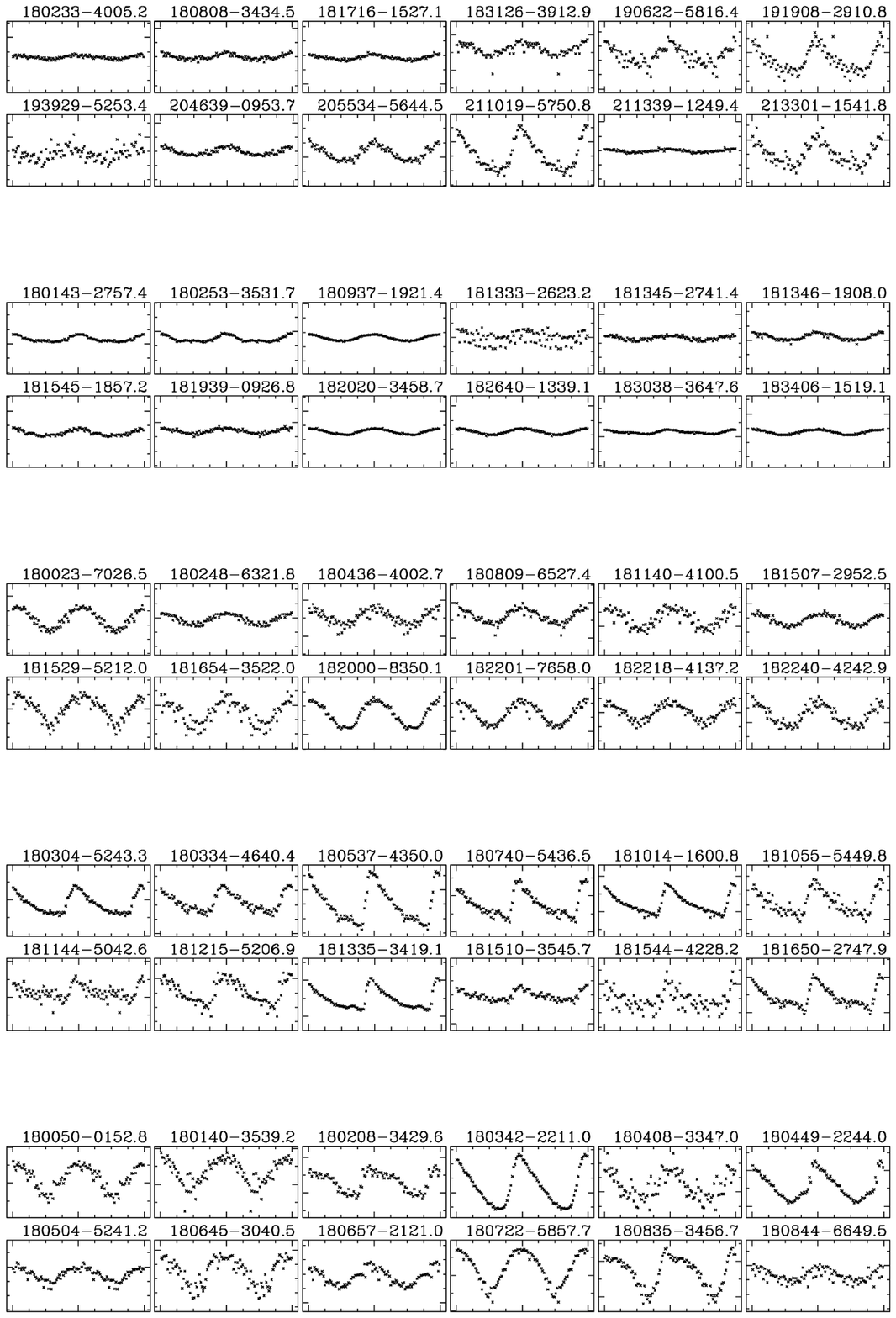}}
\end{figure}

\begin{figure}[p]
\vskip-2mm
\centerline{Stars classified as DCEP-FO}
\vskip3.5cm
\centerline{Stars classified as CW}
\vskip3.5cm
\centerline{Stars classified as M}
\vskip3.5cm
\centerline{Stars classified as MISC}
\vskip3.5cm
\vskip-15.2cm
\centerline{\includegraphics[bb=30 65 405 610, width=13cm]{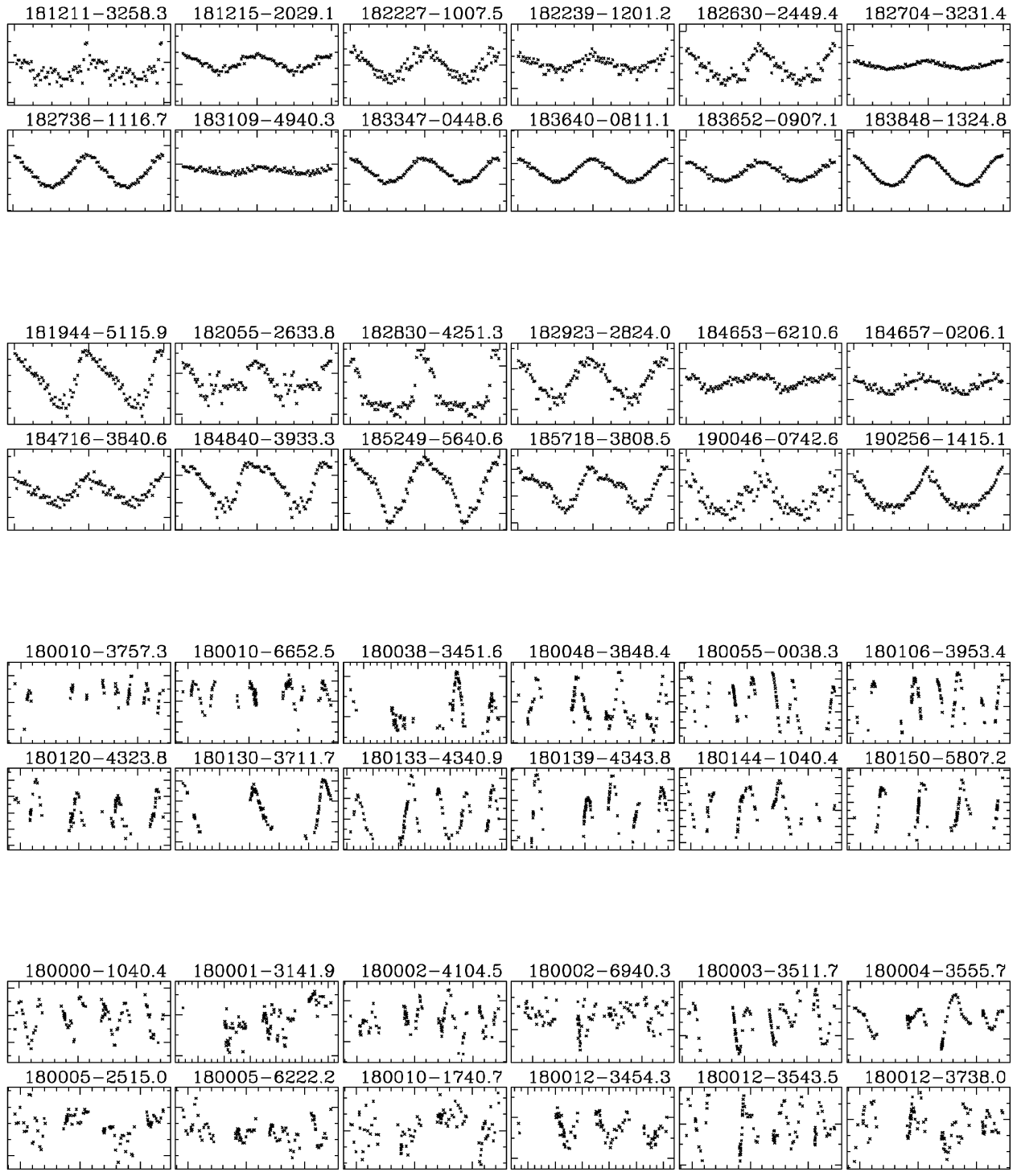}}
\end{figure}

\end{document}